\documentclass{ws-procs9x6}

\usepackage{latexsym}
\usepackage{amssymb}
\usepackage{amsmath}

\renewcommand{\rho}{\varrho}
\renewcommand{\phi}{\varphi}
\renewcommand{\epsilon}{\varepsilon}
\renewcommand{\theta}{\vartheta}
\renewcommand{\kappa}{\varkappa}   

  \newcommand{\fract}[2]{{\textstyle{\frac{#1}{#2}}}}

  \newcommand{\LL}{\left}
  \newcommand{\RR}{\right}

  \newcommand{\ssh}{\slash\!\!\!}

  \newcommand{\dd}{{\rm d}}
  
  \newcommand{\spacep}{{\bf p}}
\begin{document}
\title{Non-commutative topologically massive gauge theory }
\author{Nicola Caporaso$^{\dag}$, Luca Griguolo$^{\star}$,
Sara Pasquetti$^{\star}$  and Domenico Seminara$^{\dag}$}
\address{$^\dag$ Dipartimento di Fisica, Polo Scientifico, Universit\`a
          di Firenze; INFN Sezione di Firenze
           Via  G. Sansone 1, 50019 Sesto Fiorentino, Italy\\
          $^\star$ Dipartimento di  Fisica, Universit\`a  di Parma;
          INFN--Gruppo Collegato di Parma
          Parco Area delle Scienze 7/A, 43100 Parma, Italy}
\maketitle
\abstracts{We investigate the perturbative dynamics of noncommutative topologically massive gauge
 theories with softly broken supersymmetry. The deformed dispersion relations induced by noncommutativity
 are derived and their implications on the quantum consistency of the theory are discussed.
}



\section{Introduction}

Non-commutative quantum field theory is a fascinating theoretical
laboratory where highly non-trivial deformations of space-time
structures induce novel and unexpected dynamical effects at
quantum level. Recently they have attracted a lot of attention,
mainly due to the discovery of their relation to string/M theory
\cite{cds,sw}. In particular, Seiberg and Witten  \cite{sw}
realized that a certain class of quantum field theories on
non-commutative Minkowski space-times can be obtained as a
particular low energy limit of open strings in the presence of a
constant NS-NS $B$-field. From a purely field theoretical point of
view they appear as a peculiar non-local deformation of
conventional quantum field theory, presenting a large variety of
new phenomena not completely understood, even at perturbative
level. Four dimensional non-commutative gauge theories are in fact
afflicted by the infamous UV/IR mixing \cite{ss} that complicates
the renormalization program and it may produce tachyonic
instabilities \cite{ae}. We will not try to address these problems
in D=4\cite{VB}: our investigations will be instead concentrated on the
three dimensional, non-commutative, topologically massive
electrodynamics for a number of reasons. First of all the presence
of a single physical polarization and of an explicit
gauge-invariant mass for the photon should simplify the analysis
of the UV/IR mixing and elucidate the nature of the tachyonic
instabilities. Secondly, planar non-commutative gauge theories
with Chern-Simons terms have been proposed as effective
description of the Fractional Hall Effect \cite{sk,BarbPar}. Last
but not least, two of us\footnote{L.G. and D.S. plead guilty for
leading astray the innocent souls of N.C. and S.P. with this
project.} share with Stanley an insane passion for the ubiquitous
Chern-Simons term and its unusual dynamical properties: we hope he
will enjoy our non-commutative exercises on topologically massive
gauge theory, of which he is a Master.

\section{Non-commutative $U(1)$ Yang-Mills-Chern-Simons}

\subsection{Pure gauge model, and its symmetries}
Non-commutative\footnote{In what follows we shall only consider
the case of space-like and therefore the  constant tensor
$\theta^{\mu\nu}$ is chosen to be  $\theta\epsilon^{0\mu\nu}$,
where the index $0$ denotes the time component.} topologically
massive $U(1)$ gauge theory in three dimensions is governed by the
Lagrangian
\begin{equation} \label{ymcs_action}
    S=
    -\fract{1}{4}\int\!\! \dd^3x\; F_{\mu\nu}\star F^{\mu\nu}
   \!\! -m_g\!\!\int\!\! \dd^3x\;\varepsilon^{\lambda\mu\nu}\left(\fract{1}{2}A_{\lambda}\star\partial_{\mu}A_\nu
    +\fract{ig}{3} A_{\lambda}\star A_{\mu}\star A_\nu\right),
\end{equation}
where the field-strength is given by
\begin{equation}
F_{\mu\nu}=\partial_\mu A_\nu-\partial_\nu A_\mu-i
g[A_\mu,A_\nu]_\star
\end{equation}
and the $\star$ stands for the usual Moyal product,
\begin{equation}
    \LL(f\star g\RR)(x)
    \doteq
    \int
    \frac{\dd^2 y}{\pi\vartheta}
    \int
    \frac{\dd^2 z}{\pi\vartheta}
    f(t,y)g(t,z)
    e^{-\frac{2i}{\theta}\varepsilon_{0ij}(y-x)^i(z-x)^j}.
\end{equation}
Differently from the commutative case, which simply describes the
propagation of a free massive boson, the new non-commutative
incarnation is an interacting theory, resembling more a non-abelian
model than an abelian one\footnote{This is not surprising, because the
actual  gauge group of (\ref{ymcs_action}) can be identified with a particular
realization of $U(\infty)$.}.


\noindent This richer structure at the level of interactions is
however paid when considering the global symmetries. The constant
tensor $\theta^{\mu\nu}$ present in the definition of the Moyal
product does waste the original Lorentz invariance.[In three
dimensions, there is no Lorentz-invariant constant antisymmetric
two-tensor.] In the case of space-like non-commutativity
($\theta^{\mu\nu}\theta_{\mu\nu}<0$), the residual symmetry can be
identified with the spacial rotation $SO(2)$ and the translations.
For  time-like
non-commutativity($\theta^{\mu\nu}\theta_{\mu\nu}>0$), $SO(2)$ is
replaced by $SO(1,1)$, but the theory is not unitary\cite{gomis1}.
Finally, dealing with light-like non-commutativity
($\theta^{\mu\nu}\theta_{\mu\nu}=0$) is trickier, but one can show
that the residual group is "$\mathbb{R}$". The discrete symmetries
$(C,P,T)$ instead follow closely the known path of the commutative
case:  $C$ is conserved, while $P$ and $T$ are again broken.

\noindent
The equations of motion derived from the action \eqref{ymcs_action},
but not the action itself, are obviously gauge invariant against the
$\star-$gauge transformation
\begin{align}
    A_\mu^u(x)
    \doteq
    u(x)\star A_\mu(x)\star u^\dag(x)
    -\frac{i}{g}\partial_\mu u(x) \star u^\dag(x),
\end{align}
generated by $\star-$unitary functions $u(x)\star u^\dag(x) =
u(x)^\dag\star u(x)= 1$. The presence of the Chern--Simons term in
\eqref{ymcs_action} produces indeed a non-vanishing variation,
\begin{eqnarray}
&&\delta S=\frac{m_g}{6g^2}\int \dd^3x \,\epsilon^{\lambda\mu\nu}
(u^\dag \star \partial_\lambda u \star u^\dag \star \partial_\mu u \star
u^\dag \star \partial_\nu u)-\\&&-\frac{m_g}{2g}
    \int \dd^3x \,\epsilon^{\lambda\mu\nu}
    \partial_\lambda (u^\dag\star\partial_\mu u \star A_\nu)
=4\pi^2\LL(\frac{m_g}{g^2}\RR) w(u)
    +\text{total divergence}\nonumber
  \end{eqnarray}
where $w(u)$ is the non-commutative version of the usual
winding index. An example of transformation $u$,
for which $w(u)$ is not zero, is given by
\begin{equation}
  u_{\mathbb{P}}=[1-\mathbb{P}(x,y)]+e^{ib(t)}\mathbb{P}(x,y),
\end{equation}
where $\mathbb{P}(x,y)=2\exp(-(x^2+y^2)/|\theta|)$ is a $\star-$projector
($\mathbb{P}\star\mathbb{ P}=\mathbb{P}$) and $b(t)$ is any function
such that $b(t)\big|_{-\infty}^{\infty}= 2\pi$. In this particular case,
we find $w(u)=1$. 
Thus, as  occurs for  non-abelian topologically massive gauge
theory\cite{DJT}, the consistency of the quantum theory requires
that the mass $m_g$ is quantized according to the  relation
\begin{align}
    4\pi^2 \LL(\frac{m_g}{g^2}\RR)
    =2\pi k.
\end{align}
\subsection{The ${\mathcal{N}}=1$ supersymmetric extension}
At the perturbative level one of the most puzzling feature of
non-commutative field theory is the phenomenon of the
ultraviolet-infrared (UV/IR) mixing. The non-local nature of the
interaction, while softening the behavior at large momenta, moves
the UV divergences into the IR region. This effect
generically endangers the stability of the perturbative vacuum,
the unitarity and the infrared finiteness of the theory.

\noindent
An elegant  way to have under control these potential
problems is to consider the supersymmetric  extension of the model.
Supersymmetry improving the ultraviolet behavior of a theory will
also act, via (UV/IR) mixing, as an infrared regulator. In fact,
if  the number of supersymmetries is sufficiently large, all the
undesired divergences will disappear from the infrared region.

\noindent In three dimensions, for the case of the Yang-Mills
Chern-Simons (YMCS) system, it is enough to consider the ${
\mathcal{N}}=1$ extension of the model, whose Lagrangian is obtained by
minimally coupling a Majorana fermion to the action
(\ref{ymcs_action}),
\begin{eqnarray} \label{symcs_action}
    &&S^{\rm NC}_{\rm S-YMCS}=
    -\frac{1}{4} \int \dd^3x F_{\mu\nu}\star F^{\mu\nu}
    +\frac{1}{2} \int \dd^3x \bar \lambda \star \LL(i\ssh D-m_f\RR)\star\lambda
    +\nonumber\\&&
    -\frac{1}{2}m_g\varepsilon^{\lambda\mu\nu} \int \dd^3x A_{\lambda}\star\partial_{\mu}A_\nu
    +\frac{i}{3}g m_g \int \dd^3x\varepsilon^{\lambda\mu\nu}A_{\lambda}\star A_{\mu}\star A_\nu.
\end{eqnarray}
In eq. (\ref{symcs_action}) we have also softly broken the
supersymmetry to $\mathcal{N}=0$ by choosing different masses for the gauge
field and the Majorana fermion. This will not jeopardize the
cancellation of the infrared singularities because, in this case,
they are related just to the leading ultraviolet divergences.
Besides this breaking will provide us with a much richer and
interesting model: by taking, in fact, different limits for the
masses, we can focus our attention, for example, either on the
pure bosonic theory ($m_f\to \infty$) or on the usual
supersymmetric gauge theory ($m_f=0$, $m_g=0$) or on  the
Chern-Simons theory ($m_g\to\infty$).

\noindent Finally a remark is in order. Naively one may expect
that there is no problem with the UV/IR mixing for the  YMCS
system. In fact topologically massive commutative  gauge theories
are super-renormalizable models, that actually result  UV-finite
in perturbation theory. Thus, apparently, there is no UV
divergence to be moved in the IR region. However their finiteness
originates partly from their symmetries: the simultaneous presence
of Lorentz and gauge invariance forbids the potential linear
divergences. In the non-commutative set-up Lorentz invariance is
lost and the linear divergences will reappear as  infrared
divergences via (UV/IR) mixing. However the theory is still
UV-finite.

\section{The one-loop two-point function}
The simplest way to address the question of vacuum stability and unitarity
is to analyze the one-loop one-particle irreducible two-point function for
the gauge boson.
At the tree level, this function coincides with the commutative one since
the $\star-$product is irrelevant in the quadratic part of the action
(\ref{ymcs_action})\footnote{It holds
the following property:
$$
\int f\star g=\int fg.
$$}. Its tree level form in the Landau gauge  is in fact
\begin{equation}
\label{Gammatree}
\Gamma^{\mathrm{tree}}_{\mu\nu}(p)= \eta_{\mu\nu} p^2-p_\mu p_\nu-i m_g \epsilon_{\mu\nu\lambda}
p_\lambda.
\end{equation}
In the commutative case, when computing the one-loop correction, 
the only effect of the radiative corrections is to properly
renormalize the two transverse structures in (\ref{Gammatree}). In fact they
can be recast in the general form
\begin{equation}
\label{Gammaoneloop}
\Pi_{\mu\nu}(p)= \Pi_{e}(p)(\eta_{\mu\nu} p^2-p_\mu p_\nu)-i m_g
  \Pi_{o}(p)\epsilon_{\mu\nu\lambda}
p_\lambda.
\end{equation}
The two functions $\Pi_e$ and $\Pi_o$, computed in \cite{DJT,PisarskiRao}, govern the commutative wave-function
and the mass renormalization respectively.

\noindent This simple setting cannot be promoted to the
non-commutative case as it originates from the simultaneous
presence of
gauge and Poincar\'e invariance  which is now broken.
Once the Lorentz invariance is lost, we cannot  expect just one
wave-function ($\mathcal{Z}_{e}=1-\Pi_e$) and mass
($\mathcal{Z}_{m}=1-\Pi_o$) renormalization, since different
components of the gauge field may renormalize in different ways.
More importantly, even the transversality of the one-loop
correction to the $\Gamma^{\mathrm{tree}}_{\mu\nu}$ may be
endangered. This possibility, for example, takes place in the non
abelian gauge theory at finite temperature
\cite{PisaYaffeGross,Weldon}, where the space-time symmetries are
destroyed by the  choice of a preferred reference system, the
thermal bath.

\noindent Therefore, before proceeding, we must carefully
reexamine the Ward Identity that controls the longitudinal part of
the $\Pi_{\mu\nu}$. A tedious exercise, with the
non-commutative version of the BRST transformation, shows that in
any covariant $\xi$ gauge (and thus also in the Landau gauge) the
following Ward identity holds
\begin{eqnarray}
  p_\lambda \, \Pi^{\lambda\alpha}(p)
  =g   \Gamma_\nu(p)\left(p^\nu p^\alpha-p^2\delta^{\nu\alpha}
  -i m_g\epsilon^{\nu\alpha\beta} p_\beta
  -i \Pi^{\nu\alpha}(p) \right)
\end{eqnarray}
where   $ \Gamma_\nu $ is defined through the following vacuum
expectation $ \langle \bar c(x) [A_\nu(y), c(y)]_\star
\rangle_0\equiv i \int d^3 z \mathcal{G}(x-y-z) \Gamma_\nu (z), $
with $\mathcal{G}$ the exact ghost propagator. In the commutative
case $\Gamma_\nu $ is compelled by Lorentz invariance  to be
proportional to $p_\nu$ and the above identity entails
transversality. In the non-commutative model, there are two new
possible vectors  that can appear in the expansion of
$\Gamma_\nu$,
\begin{equation}
\tilde p^\mu=\theta^{\mu\sigma}p_\sigma\ \ \ \  \ \chi^\mu=\epsilon^{\mu\alpha\beta}\tilde p_\alpha p_\beta,
\end{equation}
and the above argument seems to break. However a detailed one-loop
analysis shows that $\Gamma^\nu$ has surprisingly  no component
along $\tilde p^\mu$ and $\chi^\mu$. Therefore the transversality
is preserved at one-loop. At higher loops, the situation is less
clear, but there are indications that this property is preserved.

\noindent
Once we have convinced ourselves that the transversality is kept, we can write the most general form
for the $\Pi_{\mu\nu}$, which is also compatible with the bosonic symmetry
\begin{eqnarray}
  \label{self}
  \Pi_{\mu\nu}=
  \Pi^{\rm e}_1 p^2 \frac{\chi_\mu \chi_\nu}{\chi^2}
  + \Pi^{\rm e}_2 \frac{\tilde p_\mu\tilde p_\nu}{\tilde p^2} p^2
  -\Pi^{\rm o} im_g \epsilon_{\mu\nu\lambda}p^\lambda
  +\bar\Pi^{\rm o}(\tilde p_\mu \chi_\nu+\tilde p_\nu \chi_\mu).
\end{eqnarray}
Actually the last tensor structure will not appear at any order in
perturbation theory because of the accidental invariance
$\theta\to -\theta$ that  $\Pi_{\mu\nu}$ possesses. This, combined
with the Bose symmetry, implies that $\bar\Pi^{\rm o}$ must be
even in $\theta $ and odd in $p$ but  such a scalar cannot be
built. We are left with
\begin{eqnarray}
  \label{self-1}
  \Pi_{\mu\nu}=
  \Pi^{\rm e}_1 p^2 \frac{\chi_\mu \chi_\nu}{\chi^2}
  + \Pi^{\rm e}_2 \frac{\tilde p_\mu\tilde p_\nu}{\tilde p^2} p^2
  -\Pi^{\rm o} im_g \epsilon_{\mu\nu\lambda}p^\lambda.
  \end{eqnarray}
At the end of the day the only effect of non-commutativity is to produce two different
wave-function renormalization: one for the component $(\tilde p\cdot A)$
and one for the component $(\chi\cdot A)$. The commutative case (\ref{Gammaoneloop}) is recovered
when $\Pi^{\rm e}_1=\Pi^{\rm e}_2$, because $\eta_{\mu\nu}-p_\mu p_\nu/p^2=
{\chi_\mu \chi_\nu}/{\chi^2}+{\tilde p_\mu\tilde p_\nu}/{\tilde p^2}$.

\noindent
Summing the general form (\ref{self-1}) of the radiative correction  to the
tree level contribution (\ref{Gammatree}) and inverting the total result, we
obtain the renormalized propagator
\begin{eqnarray}\nonumber
  G^R_{\mu\nu}(p)
  =
  \frac{1}{\sqrt{\mathcal{Z}_1 \mathcal{Z}_2}[p^2-(m^{\rm R}_g)^2] }
  \Bigg(
  \sqrt{\frac{\mathcal{Z}_2}{\mathcal{Z}_1}}
\frac{\chi_\mu \chi_\nu}{\chi^2}
  +\sqrt{\frac{\mathcal{Z}_1}{\mathcal{Z}_2}}\frac{\tilde p_\mu\tilde p_\nu}{\tilde p^2}
  + im^{\rm R}_g\epsilon_{\mu\nu\lambda}\frac{p^\lambda}{p^2} \Bigg),
\end{eqnarray}
where
\begin{eqnarray}
 \nonumber \mathcal{Z}_1=1-\Pi_1^{e}
  \ \ \
  \mathcal{Z}_2
  =&&
  1-\Pi_2^{e}
  \ \ \
  \mathcal{Z}_m=1-\Pi^{o}\ \ \
   (m^{\rm R}_g)^2=
  \frac{m_g^2 \mathcal{Z}_m^2}{\mathcal{Z}_1\mathcal{Z}_2}.
\end{eqnarray}
In next section, by looking at different features  of
$G^R_{\mu\nu}(p)$ at one-loop, we shall illustrate how the
non-commutativity affects the spectrum of the theory, its
unitarity and its vacuum stability. But for accomplishing that, we
need the explicit form of scalar functions $\Pi^{\rm e}_1$,
$\Pi^{\rm e}_2$ and $\Pi^{\rm o}$, whose evaluation is lengthy and
tedious. In the following  we shall not report on the details of
the computations, which will appear in \cite{CGPS}. The final
result is given for completeness in appendix A. Here we shall
limit ourselves to some general comments on their properties. Each
function displays two contributions, which originates respectively
from the  "planar" and  "non-planar" diagrams. The former is
identical to the commutative (non abelian) case, while the latter
carries the effects of the non-commutativity. They are both finite
and they cancel each other when $\theta$ goes to zero. This
decoupling occurs since the softly broken supersymmetric model
smoothes the effect of the UV/IR mixing. Finally  both
contributions possess a physical threshold  at $p^2=4 m^2_g$, but
two unphysical threshold at $p^2=0$ and $p^2=m^2_g$. The last
feature will complicate our future analysis.

\section{Dispersion relation and the stability of the vacuum}
The spectrum of the non-commutative Yang-Mills Chern-Simons system
is entirely encoded in the poles of the above propagator\footnote{Similar investigations
has been performed in \cite{BarbPar}}. Firstly
it contains an unphysical pole at $p^2=0$, which describes the
longitudinal degree of freedom still propagating in any covariant
gauge. Secondly, it contains the relevant physical pole at
\begin{equation}
\label{pole1}
p^2=(m_g^{\rm R})^2(p,\tilde  p)= \frac{m_g^2 \mathcal{Z}_m^2(p,\tilde p)}{\mathcal{Z}_1(p,\tilde p)
\mathcal{Z}_2(p,\tilde p)},
\end{equation}
which represents the effect of the radiative corrections on the tree level pole at $p^2=m_g^2$.
Since the Lorentz invariance is broken, eq. (\ref{pole1}) does not depend only on $p^2$ but also
on the new invariant $\tilde p^2$, which  is simply proportional to the
euclidean norm of the spacial momentum for the case of space-like non-commutativity. Therefore
the pole condition (\ref{pole1}) should not be thought as an equation for evaluating the radiative corrected
mass, but rather as an equation that determines the energy of the excitation in terms of  its
momentum, namely the dispersion relation. In a relativistic theory, this question is pointless
because the functional form of the dispersion relation is fixed by the Poincar\'e symmetry.

\noindent
The simplest way to solve  eq. (\ref{pole1}) is to proceed perturbatively. At the lowest order
in $\displaystyle{\left(\frac{g^2}{m_g}\right)}$ we have:
\begin{eqnarray}
\label{disper1} \!\!\!\!E^2  = \vec{p}^2+
m_g^2\left[1-\left(\frac{g^2}{m_g}\right)\bigg(2\Pi^o(p,\tilde
p)-\Pi^e_1(p,\tilde p)- \Pi^e_2(p,\tilde p)\bigg)
\bigg|_{p^2=m_g^2}\right]\!\!,
\end{eqnarray}
where we have factored out the dependence of the  one-loop $\Pi$s
on the coupling constant. In order that eq. (\ref{disper1})
provide a reasonable dispersion relation for a stable physical
excitation, two criteria must be met: (a) it has to be gauge
invariant; (b) it has to be real. These two requirements are far
from being manifest, since the explicit form of  the $\Pi$s is
plagued by many complex contributions (see appendix A) coming from
the unphysical thresholds at $p^2=0$ and $p^2=m_g^2$ and moreover
our perturbative computation has been performed in the Landau
gauge.

\noindent The first point can be easily clarified by evaluating
the combination $2\Pi^o(p,\tilde p) -\Pi^e_1(p,\tilde p)-
\Pi^e_2(p,\tilde p)$ at the threshold $p^2=m^2_g$. A series of
unexpected cancellations occur and the final result is completely
real. This apparent miracle is just a signal that unitarity is
preserved. The above combination can be in fact reinterpreted as
the  S-matrix element describing the transition from one particle
state to one particle state\cite{PisarskiRao}. Thus, if unitarity
is not violated, this element must be free from  unphysical cuts.

\noindent
The interpretation of the combination $2\Pi^o(p,\tilde p)-\Pi^e_1(p,\tilde p)- \Pi^e_2(p,\tilde p)$ for
$p^2=m^2_g$ as an element of the S-matrix also solves the second puzzle. In fact we know that S-matrix elements
are gauge invariant. An alternative proof can be also given by means of the Nielsen identity.

\noindent
For space-like non-commutativity,  the  explicit form of the gauge boson dispersion relation (\ref{disper1})
reads
\begin{eqnarray}
\label{disper2}
   {}&&{}\!\!\!\!\frac{E^2_{g}}{m_g^2}\!
    =\!
    1\!\!
    +\!\!\frac{\spacep^2}{m_g^2}\!
    +\!\!
    \frac{g^2}{8\pi m_g}
    \Bigg\{\!\!
    (1+2\mu)^2\!\!\!
    \LL(\!
    \int_0^1\!\! dx
    \frac{e^{-\xi\sqrt{\mu^2-x+x^2}}}{\sqrt{\mu^2-x+x^2}}
    -\log\!\LL(\frac{2|\mu|+1}{2|\mu| -1}\!\RR)\!\!
    +\frac{1}{|\mu|}\!\RR)\nonumber
    \\
    &&\!\!\!\!
    -\!27\!\!
    \LL(
    \int_0^1\!\! dx
    \frac{e^{-\xi\sqrt{1-x+x^2}}}{\sqrt{1-x+x^2}}
    -\log 3
    \RR)
    {}{}\!\!\!\!\!
    +4\!\LL(\frac{e^{-|\mu| \xi}-e^{-\xi}}{\xi}-\frac{1+4\mu+4|\mu|}{4|\mu|}\RR)\!\!
    \Bigg\},\!\!\!\!\!\!\!\!
  \end{eqnarray}
where we have introduces the dimensionless variables
$\xi=m_g\tilde p$ and $\mu=m_f/m_g$ for  convenience. In this
dispersion relation we can distinguish essentially three terms:
the first bracket contains the fermion contribution, the second
parenthesis collects instead the gauge contribution, while the
last piece is the remnant of the UV/IR mixing. This expression in
fact finite in the infrared region $\xi\to 0$ for finite $\mu$. In
the mere supersymmetric case $\mu=1$ eq. (\ref{disper2}) dramatically
simplifies and we are left  just with the bosonic contribution,
but with a different coefficient: $-18$ instead of $-27$.

 \noindent If $\mu\to \infty$, i.e. if we approach the pure
Yang-Mills Chern-Simons system, the UV/IR mixing will rise again. In fact the
last bracket will produce an infrared divergent term of the form $
-4 e^{-\xi}/\xi$. The rising of this  negative divergent
contribution at small $\xi$ for   sufficiently large $\mu$ will
always make the square of the energy negative in a certain region
of the spacial momenta (see fig. 1).
\begin{figure}[hbtt]
\centerline{\epsfxsize=4.5in\epsfysize=2.6in\epsfbox{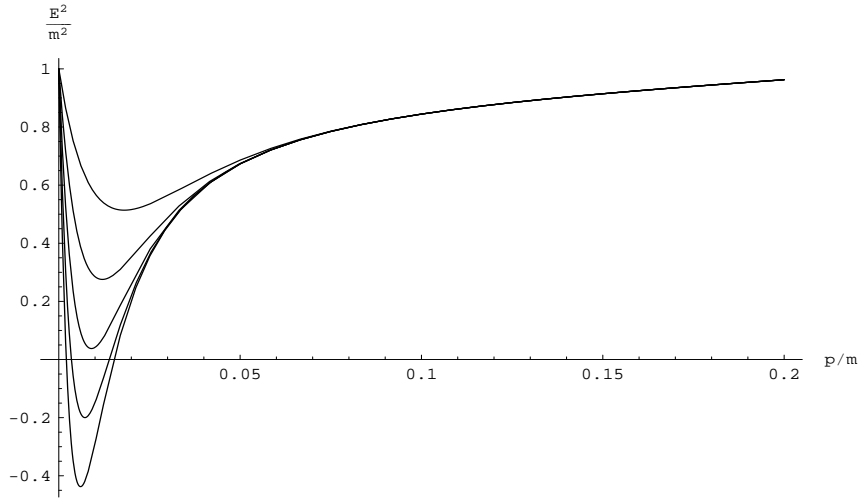}}
\caption{The dispersion relation of the boson for small spacial
momenta, when $g^2/m_g=0.1$ $m^2_g\theta=1$ and $\mu=100\to 300 $.
\label{pippo}}
\end{figure}
\noindent In other words, the massive excitation becomes a tachyon
and the perturbative vacuum is no longer stable. Varying the other
two parameters $g^2/m_g$ and $m^2_g\theta$ will not affect the
picture: it will only change the specific value of $\mu$ at which
the tachyon will appear. Thus, when we reach the critical value of
$\mu$, we must  to resort with non-perturbative tecniques to
select the new vacuum. At the moment, the nature of this new
vacuum is only matter of speculation\cite{BarbPar}. One may conjecture  that the
transition tuned by the tachyonic mode will lead the system to a
sort of stripe phase analogous to that proposed by Gubser and
Sondhi \cite{GS} for $\phi^4$. But this possibility is quite
problematic: a non translationally invariant vacuum would mean a
dynamical breaking of the gauge invariance and this could endanger
the consistency of the entire theory. Recall that, for a
non-commutative gauge theory, space-time translations are in fact
a subset of the gauge transformations.

\noindent A less speculative point of view, but nevertheless very
intriguing, is to suppose that the tachyonic mode will drive the
Yang-Mills Chern-Simons system through a phase transition similar
to the one speculated by Cornwall \cite{Cornwall} for the non
abelian model, in the commutative case, in the large $N-$limit. We
must recall in fact that there is a great similarity between
non-commutative gauge theories and gauge theories at large
N\cite{Szabo}.

\noindent The fate of the perturbative vacuum should be discussed,
of course, at non-perturbative level: a possibility is by
employing the matrix model representation of the theory and it
will be the object of future investigations.

\appendix
\section{Analytic expression of  the different $\Pi$s}
If we introduce the following basic integrals ($k=-1,0,1$)
\begin{equation}
\!
T^{\textrm{(np)}}_{k}(\mu_1, \mu_2)\! =\!\frac{\partial^{k+1}}{\partial \xi^{k+1}}
\left(\int_0^1 d x
\frac{(-1)^{k+1} e^{-\xi\sqrt{x(x-1)\eta^2+x\mu_1^2+(1-x)\mu_2^2}}}{\sqrt{x(x-1)\eta^2+x
\mu_1^2+(1-x)\mu_2^2}}\;\right)
\end{equation}
 and the dimensionless variable $\eta^2=p^2/m^2_g$, the non-planar contributions for the different
 $\Pi$s are given by\\
{\it \bf Gluon Sector:}
\begin{eqnarray}
\Pi^{\rm e}_{1,\rm np}\!\!\! =&&\!\!\!\!
  -\nonumber\frac{g^2}{8\pi m_g}\Bigg[
  \Bigg(\frac{9(4-\eta^2)}{4}T^{\textrm{(np)}}_{-1}(1,1)
  -\frac{(4-5\eta^2+\eta^4)}{\eta^2 \xi}T^{\textrm{(np)}}_{0}(1,1)\Bigg)
  \\ \nonumber
  &&\!\!\!\!
  +\Bigg(\frac{ 5(1-\eta^2)^2}{2\eta^2}T^{\textrm{(np)}}_{-1}(1,0)
  -\frac{(6-2\eta^2)}{\xi}T^{\textrm{(np)}}_{0}(1,0)\Bigg)\!\!-\!\!
  \Bigg( \frac{\eta^2}{4} T^{\textrm{(np)}}_{-1}(0,0)- \\
  &&\!\!\!\!
   -\frac{(1-\eta^2)}{\xi} T^{\textrm{(np)}}_{0}(0,0) \Bigg) +\frac{(9-4\eta^2)e^{-\xi}-(5-4\eta^2)}{\eta^2\xi}
  \Bigg]\!\!\!
\end{eqnarray}
\begin{eqnarray}
  \Pi^{\rm e}_{2,\rm np}\!\!\!
  =\nonumber&&\!\!\!\!
  -\frac{g^2}{8\pi m_g}
  \Bigg[\Bigg(\frac{9(4-\eta^2)}{4}T^{\textrm{(np)}}_{-1}(1,1)
  +\frac{(4-5\eta^2+\eta^4)}{\eta^2 \xi}T^{\textrm{(np)}}_{1}(1,1)\Bigg)
  \\\nonumber&&\!\!\!\!
  +\Bigg( \frac{ 5(1-\eta^2)^2}{2\eta^2}T^{\textrm{(np)}}_{-1}(1,0)
  +\frac{(6-2\eta^2)}{\xi}T^{\textrm{(np)}}_{1}(1,1)
  \Bigg)\!\!-\!\!
  \Bigg(\frac{\eta^2}{4} T^{\textrm{(np)}}_{-1}(0,0)+
  \\&&\!\!\!\!+
  \frac{(1-\eta^2)}{\xi} T^{\textrm{(np)}}_{1}(0,0) \Bigg)
  +\frac{(5-4\eta^2)-(9-4\eta^2)e^{-\xi} }{\eta^2\xi}\Bigg]\!\!\!
\end{eqnarray}
\begin{eqnarray}
  \Pi^{\rm o}_{\rm np}\!\!\!
  =&&\!\!\!\!
  -\nonumber\frac{g^2}{16\pi m_g}
  \Bigg[\Bigg(\frac{3}{2}\eta^4-2\eta^2\Bigg)T^{\textrm{(np)}}_{-1}(0,0)
  -\Bigg(3\eta^2-\frac{3}{2}\eta^4+12\Bigg)T^{\textrm{(np)}}_{-1}(1,1)
  \\&&\!\!\!\!
  -\frac{(1-\eta^2)^2(1+3\eta^2)}{\eta^2\xi}T^{\textrm{(np)}}_{-1}(1,0)
  +2\frac{(1-2\eta^2)}{\eta^2}\frac{1 - {}{e^{-\xi }}}{\xi }\Bigg]
\end{eqnarray}
{\it\bf Fermion Sector:}
\begin{eqnarray}
     \Pi^{\rm e}_{\rm 1,np}
    \!\!=&&\!\!\!-
    \frac{g^2}{2\pi m_g\eta^2}
     \Bigg(T^{\mathrm{(np)}}_{1}(\mu,\mu)-\mu^2T^{\mathrm{(np)}}_{-1}(\mu,\mu)\Bigg)
   ,\ \Pi^{\rm 0}_{\rm np}\!\!
    =\!\!
    -\frac{g^2\mu }{4\pi m_g}T^{\mathrm{(np)}}_{-1}(\mu,\mu),\nonumber
    \\
    \Pi^{\rm e}_{\rm 2,np}
    \!\!=&&\!\!\!{-}
    \frac{g^2}{2\pi m_g\eta^2}
    \Bigg(\mu^2T^{\mathrm{(np)}}_{-1}(\mu,\mu)+\frac{1}{\xi} T^{\mathrm{(np)}}_{0}(\mu,\mu)
    \Bigg).
\end{eqnarray}
The planar contribution are identical to that of Pisarski and Rao \cite{PisarskiRao}.

\end{document}